\documentclass[twocolumn,prl]{revtex4-1}
\usepackage{color}
\usepackage{amsmath}
\usepackage{amssymb}
\usepackage{graphicx}
\usepackage[unicode=true,pdfusetitle,
 bookmarks=true,bookmarksnumbered=false,bookmarksopen=false,
 breaklinks=false,pdfborder={0 0 0},pdfborderstyle={},backref=false,colorlinks=true]
 {hyperref}
\hypersetup{
 colorlinks,linkcolor=red,citecolor=blue}

\makeatletter

\usepackage{amsfonts}

\makeatother

\begin{document}

\title{Synthetic gauge field in a single optomechanical resonator}

\author{Yuan Chen$^{1,2,\dagger}$, Yan-Lei Zhang$^{1,2,\dagger}$, Zhen
Shen$^{1,2,\dagger}$, Chang-Ling Zou$^{1,2}$, Guang-Can Guo$^{1,2}$,
and Chun-Hua Dong$^{1,2}$}

\affiliation{$^{1}$CAS Key Laboratory of Quantum Information, University of Science
and Technology of China, Hefei 230026, P. R. China.}

\affiliation{$^{2}$CAS Center For Excellence in Quantum Information and Quantum
Physics, University of Science and Technology of China, Hefei, Anhui
230026, P. R. China.}

\thanks{$^{\dagger}$These authors contributed equally to this work.}

\date{\today}

\maketitle
\textbf{Synthetic gauge fields have recently emerged }\cite{Aidelsburger2018,Hey2018}\textbf{,
arising in the context of quantum simulations, topological matter,
and the protected transportation of excitations against defects. For
example, an ultracold atom experiences a light-induced effective magnetic
field when tunnelling in an optical lattice }\cite{Goldman2014,Goldman2016}\textbf{,
and offering a platform to simulate the quantum Hall effect and topological
insulators. Similarly, the magnetic field associated with photon transport
between sites has been demonstrated in a coupled resonator array }\cite{Hafezi2013,Mittal2018}\textbf{.
Here, we report the first experimental demonstration of a synthetic
gauge field in the virtual dimension of bosonic modes in a single
optomechanical resonator. By employing degenerate clockwise (CW) and
counter-clockwise (CCW) optical modes and a mechanical mode, a controllable
synthetic gauge field is realized by tuning the phase of the driving
lasers. The non-reciprocal conversion between the three modes is realized
for different synthetic magnetic fluxes. As a proof-of-principle demonstration,
we also show the dynamics of the system under a fast-varying synthetic
gauge field. Our demonstration not only provides a versatile and controllable
platform for studying synthetic gauge fields in high dimensions but
also enables an exploration of ultra-fast gauge field tuning with
a large dynamic range, which is restricted for a magnetic field.}

\begin{figure}
\includegraphics[clip,width=1\columnwidth]{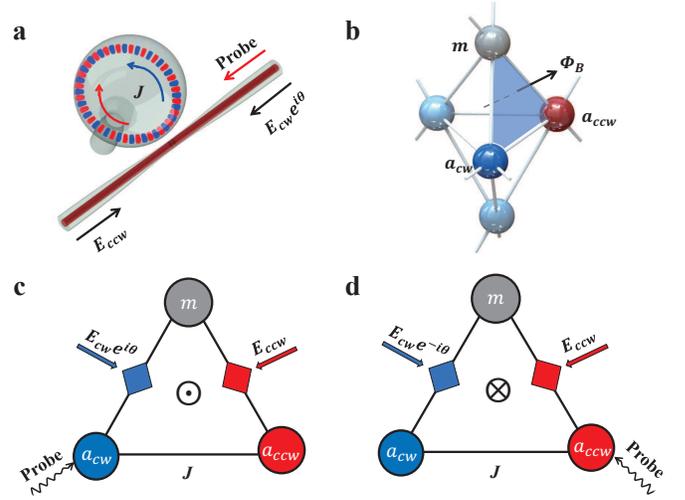}\caption{\textbf{Schematic illustration of a synthetic gauge field in an optomechanical
resonator.} \textbf{a} Two phase-related strong opposite propagating
driving fields enhance the optomechanical coupling between a mechanical
mode and the coupled optical modes inside a microcavity. $E_{\mathrm{cw\,(ccw)}}$
are the amplitudes of the clockwise (CW) and counter-clockwise (CCW)
drives, and $\theta$ is the phase difference between them. The CW
and CCW modes are coupled through backscattering in the microresonator
with the modal coupling strength of J. \textbf{b} Schematic of the
CW and CCW optical modes coupling with the radial breathing mode.
The synthetic magnetic flux $\Phi_{B}$ arises from the phase-related
driving fields. $a_{\mathrm{cw}}$, $a_{\mathrm{ccw}}$, and $m$
are the bosonic operators for the optical and mechanical modes, respectively.
\textbf{c} Energy diagram for the mode conversion between the CW ,
and CCW optical modes and the breathing mode, where the probe is transferred
from the CW (\textbf{c}) or CCW (\textbf{d}) direction. The accumulated
phase in the CW to CCW optical mode conversion is reversed for the
opposite propagating probe direction, which results in the non-reciprocal
power transmission of the optical probe field.}

\label{Fig1}
\end{figure}

When charged particles move around in a closed loop, the enclosed
magnetic flux $\Phi_{B}$ induces a phase that leads to interesting
topological phenomena in condensed matter physics. Non-trivial topological
phases and topologically protected quantum states have been extensively
explored for fundamental studies and quantum information processing.
For uncharged particles or bosonic excitations, the essential effect
of a gauge field can also be realized through a geometric phase \cite{Berry_1984},
where a particle or excitation acquires a path-dependent phase by
a carefully engineered Hamiltonian. Such a synthetic gauge field enables
the simulation of quantum many-body physics with unprecedented precision
and unconventional control of bosons \cite{Aidelsburger2018,Hey2018}.
Thus, synthetic gauge fields have aroused tremendous research interest
recently and have been realized in various systems, including ultracold
atoms \cite{Goldman2014,Goldman2016}, optical photons \cite{Hafezi2013,Mittal2018,Ozawa2019},
phonons \cite{Ma2019} and other bosonic quasi-particles \cite{Anderson2016,Roushan2017}.

A synthetic gauge field is commonly realized in real space, where
the excitation hops between sites and the phase accumulated for a
closed loop path is not trivial. For photons, it has been realized
in an artificial photonic microcavity array with engineered spatially
dependent couplings between sites \cite{Hafezi2013,Mittal2018,Ozawa2019},
while the hopping phase determined by the structural parameters is
fixed and requires nanofabrication. Recently, a reconfigurable synthetic
gauge field was proposed and realized in coupled resonators by employing
parametric interactions \cite{Estep_2014,Schmidt_2015,Fang_2017,Bernier_2017,Hey2018}.
For example, a closed loop of sites with a synthetic gauge field was
realized by two coupled optomechanical resonators \cite{Fang_2017},
where the appropriate spatially dependent hopping phases between sites
were realized through external drives.

\begin{figure*}
\centerline{\includegraphics[clip,width=1\textwidth]{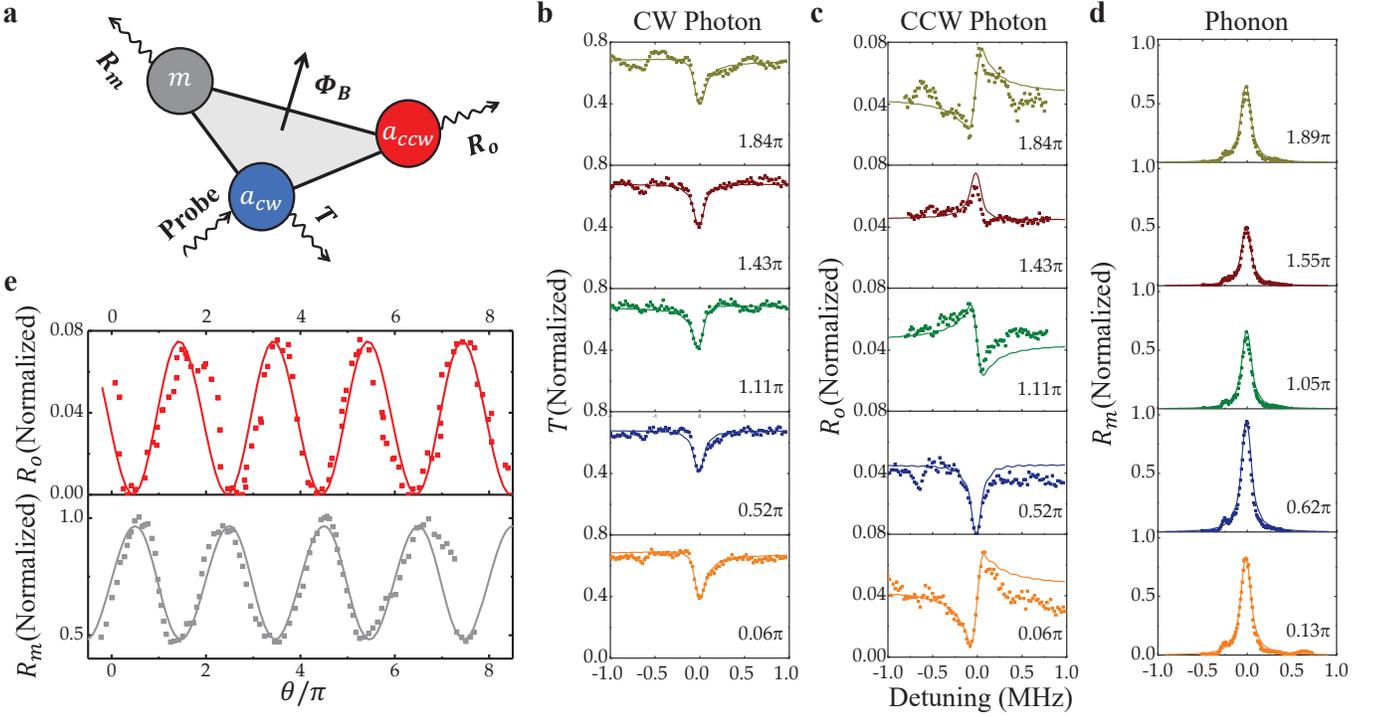}}\caption{\textbf{The typical photon-photon conversion and photon-phonon conversion
modulated by the synthetic magnetic flux.} \textbf{a} Schematic of
the mode conversions of the CW optical mode, CCW optical mode, and
mechanical mode modulated by the synthetic magnetic flux. \textbf{b},\textbf{c}
The spectra of the CW emission power ($T$) and the CW to CCW optical
mode conversion ($R_{o}$) with the synthetic magnetic flux at $\theta=0.06\pi,0.52\pi,1.11\pi,1.43\pi,1.84\pi$.
The incident driving powers are 3.7 mW (CW) and 1.6 mw (CCW). The
dots are the experimental results. The solid lines are the results
of calculations using the parameters $\omega_{m}/2\pi=98.72$ MHz,
$\kappa_{0}/2\pi=5$ MHz, $\kappa_{e}/2\pi=36$ MHz, $\gamma_{m}/2\pi=92$
kHz, $J/2\pi=5.3$ MHz, $G_{\mathrm{cw}}/2\pi=0.6$ MHz, and $G_{\mathrm{ccw}}/2\pi=0.4$
MHz, respectively. The relative optical emission power spectra of
$T$ and $R_{o}$ are calculated at 10th $\mu s$. \textbf{d} The
photon-phonon conversion ($R_{m}$) with the synthetic magnetic flux
at $\theta=0.13\pi,0.62\pi,1.05\pi,1.55\pi,1.89\pi$. The relative
phonon power density is measured at 12th $\mu s$. The corresponding
parameters are $G_{\mathrm{cw}}/2\pi=0.53$ MHz, and $G_{\mathbf{\mathrm{ccw}}}/2\pi=0.38$
MHz, respectively. The optomechanical coupling rate of the read laser
is $G_{\mathrm{cw}}^{Read}/2\pi=0.21$ MHz. The other parameters are
the same as in \textbf{b}. \textbf{e} $R_{o}$ and $R_{m}$ at $\delta=0$
are related to the synthetic magnetic flux. $R_{o}$ is nearly maximum
when $R_{m}$ is minimum. The red dots and grey dots are the experimental
results of the CW to CCW optical mode conversion and photon-phonon
conversion, respectively. The solid red line and solid grey line are
the simulation results of these conversions, respectively. The probe
propagates in the CW direction. }

\label{Fig2}
\end{figure*}

In this Letter, a synthetic gauge field is realized in a virtual dimension
by exploring the natural optical and mechanical modes in a single
spherical microresonator without the requirement of fabricating an
array of identical microstructures. The effective magnetic flux can
be precisely controlled by changing the phases of the external driving
lasers. Additionally, such a synthetic gauge field can be arbitrarily
tuned, offering the opportunity to study time-dependent gauge field
dynamics. Benefiting from the great progress achieved in the coherent
optomechanical interaction in microcavities and also the coherent
nonlinear optical effects, our demonstration of the synthetic gauge
field can be scaled up to larger dimensions but with full controllability,
which is not available in real-space lattices. Therefore, exotic topological
photonics and non-reciprocal quantum frequency conversion can be readily
realized in a single microresonator.

\begin{figure}[tp]
\centerline{\includegraphics[clip,width=1\columnwidth]{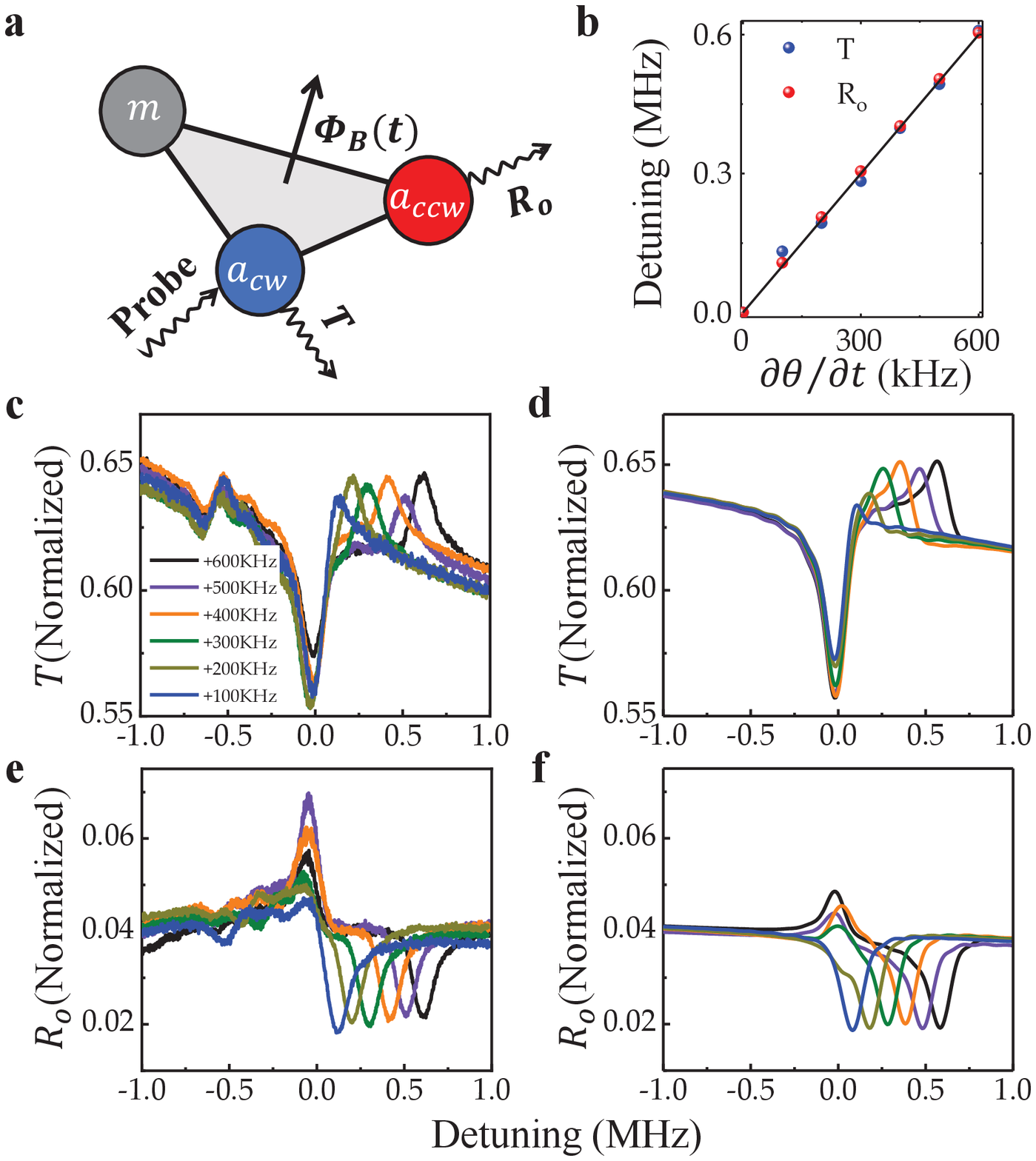}}\caption{\textbf{a} Schematic of the non-reciprocal emission power spectra
of $T$ and $R_{o}$ under a dynamic synthetic magnetic flux. \textbf{b}
Frequency detuning of $T$ and $R_{o}$ with error bars under a dynamic
synthetic magnetic flux. \textbf{c}-\textbf{f} The non-reciprocal
emission power spectra of $T$ and $R_{o}$ with a dynamic synthetic
magnetic flux $\partial\theta\left(t\right)/\partial t$ from $+100kHz$
to $+600kHz$. The experimental results are shown in \textbf{c} and
\textbf{e}, and the corresponding simulation results are shown in
\textbf{d} and \textbf{f}, respectively. The corresponding parameters
are $G_{\mathrm{cw}}/2\pi=0.69$ MHz and $G_{\mathbf{\mathrm{ccw}}}/2\pi=0.31$
MHz. The other parameters are the same as in 2\textbf{b}. }

\label{Fig3}
\end{figure}

The synthetic gauge field can be realized in a single optomechanical
system, as illustrated in Fig.$\,$\ref{Fig1}a. A single silica microsphere
supports massive high-quality optical whispering gallery modes (WGMs)
and mechanical modes \cite{Kippenberg_2007,Park_2009}, by which the
coherent conversion between the modes has been extensively studied
in the context of optomechanics \cite{Weis2010,Safavi_Naeini_2011,Dong2012,Zhang_2017}.
Therefore, by exploiting those modes as virtual dimensions, we can
study an interesting model of bosonic lattices, as depicted in Fig.$\,$\ref{Fig1}b,
with each vertex node for the bosonic mode serving as a site and the
edges corresponding to coherent hops between sites. For modes with
non-degenerate frequencies, the coherent conversion between the modes
can be stimulated by an external driving laser, which compensates
for the energy difference between the modes. In the lattice, we can
find the simplest closed loop of a triangle plaquette (blue surface
in Fig.$\,$\ref{Fig1}b), which consists of two optical modes and
one mechanical mode. In a microsphere made of a non-magnetic material,
the clockwise (CW) and counter-clockwise (CCW) optical modes are degenerate.
However, the surface roughness and material defects induce backscattering
of the photons \cite{Kippenberg_2002}, as described by an optical
modal coupling strength $J$ (Fig.$\,$\ref{Fig1}a). Both the CW
and CCW optical modes can optomechanically couple with the mechanical
mode, while the CW (or CCW) photons can only couple with the phonon
when there is an off-resonant intracavity field along the CW (or CCW)
direction due to the conservation of momentum \cite{Shen_2016}. By
introducing driving lasers in both the CW and CCW directions, coherent
coupling between the optical modes and mechanical mode can be realized,
with the hopping strength being proportional to the drives $E_{\mathrm{cw}}e^{i\left(\theta+\theta_{0}\right)}$
and $E_{\mathrm{ccw}}e^{i\theta_{0}}$. Here, $E_{\mathrm{cw(ccw)}}$
is the amplitude of the drive, and $\theta$ is the phase difference
between them. We know that the synthetic gauge field is from the phase
difference, therefore we assume the common phase factor $\theta_{0}=0$
for convenience.

Figures$\,$\ref{Fig1}c-d provide a detailed model of the triangle
plaquette, which is described by the Hamiltonian (see Supplementary
Material for details)
\begin{equation}
H=Ja_{\mathrm{cw}}a_{\mathrm{ccw}}^{\dagger}+G_{\mathrm{cw}}e^{i\theta}ma_{\mathrm{cw}}^{\dagger}+G_{\mathrm{ccw}}ma_{\mathrm{ccw}}^{\dagger}+\mathrm{H.c.},
\end{equation}
where $m$, $a_{\mathrm{cw}}$ and $a_{\mathrm{ccw}}$ are the bosonic
operators for the mechanical and optical modes, respectively. $G_{\mathrm{cw(ccw)}}\propto E_{\mathrm{cw(ccw)}}$
is the stimulated photon-phonon coupling strength produced by red-sideband-detuned
drives \cite{Weis2010,Safavi_Naeini_2011}. The remaining phase $\theta$
is gauge-independent and actually represents the phase gain by the
bosonic excitations when circulating the plaquette. Such a phase $\theta$
for bosonic excitations produces the Aharonov-Bohm effect for electrons,
where electrons travel along a closed loop and gain a phase $\theta=\frac{e}{\hbar c}\Phi_{B}$.
Here, $\Phi_{B}$ is the magnetic flux enclosed by the loop, and $e$,
$\hbar$ and $c$ are the electron charge, reduced Plank's constant
and the light velocity, respectively. Therefore, non-trivial effective
gauge fields are synthesized for photons and phonons, while the flux
$\Phi_{B}$ can be controlled optically.

In our experiments, we fabricate a microsphere with a diameter of
approximately $32\mu\mathrm{m}$, which supports degenerated high-quality-factor
WGMs near $780\,\mathrm{nm}$ (intrinsic optical linewidth of $\kappa_{0}/2\pi=5\:\mathrm{MHz}$,
and a backscattering-induced modal coupling rate of $J/2\pi=5.3\mathrm{\:MHz}$).
In the same resonator, the radial breathing mechanical mode has a
frequency of $\omega_{m}/2\pi=98.72\,\mathrm{MHz}$ and an optomechanical
mode linewidth of $\gamma_{m}/2\pi=92\,\mathrm{kHz}$. In the experiments,
both drives are generated from a laser passing through acoustic-optic
modulators (AOMs), and the probe photons are generated by an electro-optic
modulator (EOM). The phase $\theta$ of the drives is precisely controlled
through the relative phase delay between the radio-frequency (RF)
signals driving the AOMs.

To verify the synthetic gauge field in our optomechanical resonator,
we send probe laser to the optical CW mode. Comparing Figs.$\,$\ref{Fig1}c
and d, the model is symmetric under parity and time-reversal symmetry,
i.e., $\theta\rightarrow-\theta$ and CW$\rightarrow$CCW. Therefore,
the system behaviour for a probe input coupled to the optical CCW
mode (Fig.$\,$\ref{Fig1}d)is equivalent to the scheme in Fig.$\,$\ref{Fig1}c
by flipping the phase of the drives from $\theta$ to $-\theta$.
Therefore, we can prove non-reciprocal transmission by studying the
probe field from the CW port with various $\theta$ in the following
experiments. Two drives and probe pulses (duration of $\tau_{p}=10\mu s$)
are sent into the system; thus, the transient bosonic excitation transportation
under the synthetic gauge field can be experimentally investigated.
As depicted in Fig.$\,$\ref{Fig2}a, the amplitudes of $a_{\mathrm{cw}}$,
$a_{\mathrm{ccw}}$ and $m$ are separately detected. The optical
outputs are instantly and simultaneously measured since their lifetimes
are relatively low compared to mechanical mode, and the CW and CCW
outputs are separated in the forward and backward directions. To probe
the mechanical excitation, we introduce another read pulse with a
duration of $5\mu s$ at the same frequency of the drives, which arrives
$2\mu s$ after the drives \cite{Dong2012}. Because of the longer
lifetime of the phonons, the excited phonons during the first pulses
can be converted into photons and measured with a time gate detection
to obtain the displacement power spectral density of the mechanical
mode (see Supplementary Materials).

Figures$\,$\ref{Fig2}b-e summarize the experimental results of the
system with a static synthetic gauge field ($\theta$). In Figs.$\,$\ref{Fig2}b-d,
the intensities of the CW and CCW photons and the phonons for different
input probe detunings are presented. For all of the spectra, we observe
a significant change in the spectral shape due to the variation of
$\theta$, with all the other experimental conditions being fixed.
Such a phase-dependent response of the coupled three-mode system manifests
the synthetic gauge field, since the conversions from the CW probe
to CCW photon and phonon modes are effectively manipulated, which
resembles the dynamics of charged particles in a magnetic field. The
CW population contains both the direct emission of the probe field
and the weak indirect conversion from other modes of the closed loop,
only the latter part of which is related to the phase. Therefore,
the spectra are slightly modified by the synthetic gauge field $\theta$.

We further summarize the derived photon and phonon populations from
the spectra in Fig.$\,$\ref{Fig2}e. By varying $\theta$ over a
range of $7\pi$, we find the mode populations of a periodic oscillation
with a period of $2\pi$, which means the phase dependence is repeatable
and further confirms the nature of the gauge field. Since the CW photons
and phonons exhibit a complementary oscillation, the conversion efficiency
between the nodes in the loop-path of the three modes can be fine-tuned.
The breaking of time-reversal symmetry is also an important consequence
of the synthetic gauge field. Therefore, the system should show different
responses when changing $\theta$ to $2n\pi-\theta$, with $n$ being
an arbitrary integer. Such breaking of the time-reversal symmetry
is obvious in Fig.$\,$\ref{Fig2}c, and the most remarkable difference
can be found at $\theta=0.5\pi$ and $2\pi-\theta$. Additionally,
according to the symmetry under $\theta\rightarrow-\theta$ and CW$\rightarrow$CCW,
for a non-zero fixed $\theta$ that corresponds to broken time-reversal
symmetry and broken parity symmetry, non-reciprocal bosonic transportation
is most significant when $\theta=0.5\pi+2n\pi$ with the probe laser
conversion from the CW direction to the CCW direction.

The observed phase dependence and broken time-reversal symmetry of
the bosonic excitation conversion in our system verify the static
synthetic gauge field. Compared with a magnetic field, the synthetic
gauge field has the advantages of a large dynamic range and ultra-fast
tuning; thus, it provides a unique route to study the dynamics under
a fast-varying gauge field \cite{Singleton_2013,Jing_2017}. As an
example, we study the system response with a linearly varying gauge
field $\Phi_{B}\left(t\right)$, as illustrated in Fig.$\,$\ref{Fig3}a.
For various synthetic gauge fields $\partial\theta\left(t\right)/\partial t$
at a few hundred kHz, we observe a peak in the spectra of $T$, which
is off-resonant with the mechanical mode. Similarly, an extra dip
can be observed in the spectra of $R_{o}$ when compared with Fig.$\,$\ref{Fig2}c,
and the detuning approximately $\Delta\approx\partial\theta\left(t\right)/\partial t$.
Figures.$\,$\ref{Fig3}d and f show the theoretically calculated
spectra, which agree with the experimental results.\textbf{ }Such
dynamics under a temporally varying $\Phi_{B}\left(t\right)$ can
be interpreted in the frequency domain, as the frequency of the CW
drive is detuned with respect to the CCW drive by $\partial\theta\left(t\right)/\partial t$.
Therefore, the induced CW photon-phonon conversion is slightly shifted
from the resonance by $\partial\theta\left(t\right)/\partial t$ and
introduces a modification to the spectra at the detuning. In Fig.$\,$\ref{Fig3}b,
the extracted frequency of the varying gauge-field-induced peak or
dip is plotted, which shows excellent agreement with $\partial\theta\left(t\right)/\partial t$.
The results demonstrate the potential of our synthetic gauge field
to study more complex time-dependent gauge fields \cite{Singleton_2013,Jing_2017},
as an arbitrary adjustable $\Phi_{B}\left(t\right)$ can be realized
in our system by employing arbitrary wave generators for the RF AOM
inputs.

By exploiting the virtual photon and phonon degrees of freedom (DOFs)
in an optomechanical resonator, coherent coupling between the modes
can be achieved by external driving lasers, without requiring photonic
and phononic device fabrication. Beyond the simplest three-mode model,
our demonstration can be scaled to more modes in such a single resonator.
Benefiting from the enhanced nonlinear optical effects due to the
high quality factor and small mode volume, such as four-wave mixing
\cite{Gaeta2019} and Brillouin scattering \cite{Dong2015}, we could
generalize the synthetic gauge field to higher virtual dimensions.
Therefore, a single resonator is sufficient for studying interesting
topological physics in high dimensions and is feasible for manipulating
photons with unprecedented abilities. Additionally, nonlinear optical
and optomechanical effects enable the generation of entangled bosonic
excitations, which has potential for realizing topologically protected
quantum squeezing \cite{Peano2016}, entanglement \cite{Blanco-Redondo568}
and one-way quantum steering \cite{Cavalcanti2016}.

\vbox{}

\noindent\textbf{Acknowledgments}\\ The authors thank Xiang Xi for
discussions. This work was supported by the National Key Research
and Development (R\&D) Program of China (grant 2016YFA0301303), the
National Natural Science Foundation of China (grants 11722436 and
11704370, 11874342, 61805229), and Anhui Initiative in Quantum Information
Technologies (grant AHY130200). This work was partially carried out
at the USTC Center for Micro and Nanoscale Research and Fabrication.

\vbox{}

\noindent\textbf{Author contributions}\\ Y. C., Y.-L. Z. and Z.S.
contributed equally to this work. Y.-L. Z., Z.S. C.-L.Z. and C.-H.
D. conceived and designed the experiment. Y.C., Z.S. and C.H.D. prepared
the samples, built the experimental setup and carried out experiment
measurements. Y.-L.Z. and C.-L.Z. provided theoretical support and
analysed the data. C.-H.D. and C.-L.Z. wrote the manuscript with inputs
from all authors. C.-H.D, C.-L.Z. and G.-C.G. supervised the project.
All authors contributed extensively to the work presented in this
paper\emph{.}

\vbox{}

\noindent\textbf{Additional information}\\ Supplementary information
is available in the online version of the paper. Reprints and permissions
information is available online at www.nature.com/reprints. Correspondence
and requests for materials should be addressed to C.-L.Z (clzou321@ustc.edu.cn)
or C.-H.D. (chunhua@ustc.edu.cn).

\vbox{}

\noindent\textbf{Competing financial interests}\\The authors declare
no competing interests.


%

\end{document}